# *d*-wave superconductivity as a model for diborides apart MgB₂

Evgeny F. Talantsev[1,2,*]

[1]M.N. Miheev Institute of Metal Physics, Ural Branch, Russian Academy of Sciences, 18, S. Kovalevskoy St., Ekaterinburg, 620108, Russia

[2]NANOTECH Centre, Ural Federal University, 19 Mira St., Ekaterinburg, 620002, Russia

*corresponding author's E-mail: evgeny.f.talantsev@gmail.com

**Abstract**

Recently, Pei *et al*. (*arXiv*2105.13250) reported that ambient pressure $\beta$-MoB₂ (space group: $R\bar{3}m$) exhibits a phase transition to $\alpha$-MoB₂ (space group: $P6/mmm$) at pressure $P \sim 70$ GPa and this high-pressure phase is a high-temperature superconductor exhibited $T_c = 32\ K$ at $P\sim110$ GPa. Despite $\alpha$-MoB₂ has the same crystalline structure as ambient pressure MgB₂ and the $T_c$'s of $\alpha$-MoB₂ and MgB₂ are very close, the first principles calculations showed that in $\alpha$-MoB₂ the states near the Fermi level, $\varepsilon_F$, are dominated by the *d*-electrons of Mo atoms, while in MgB₂ the *p*-orbitals of boron atomic sheets dominantly contribute to the states near the $\varepsilon_F$. More recently, Hire *et al*. (*arXiv*2212.14869) reported that the $P6/mmm$-phase can be stabilized at ambient pressure in Nb$_{1-x}$Mo$_x$B₂ solid solutions, and these ternary alloys exhibit $T_c \sim 8\ K$. In addition, Pei *et al*. (*Sci. China-Phys. Mech. Astron*. **65**, 287412 (2022)) showed that compressed WB₂ exhibits $T_c \sim 15\ K$ at $P\sim121$ GPa. Here, we analyzed experimental data reported for *P6/mmm*-phases of Nb$_{1-x}$Mo$_x$B₂ (x = 0.25; 1.0) and highly-compressed WB₂, and showed that these three phases exhibit *d*-wave superconductivity. We deduced $\frac{2\Delta_m(0)}{k_B T_c} = 4.1 \pm 0.2$ for $\alpha$-MoB₂, $\frac{2\Delta_m(0)}{k_B T_c} = 5.3 \pm 0.1$ for Nb$_{0.75}$Mo$_{0.25}$B₂, and $\frac{2\Delta_m(0)}{k_B T_c} = 4.9 \pm 0.2$ for WB₂. We found that Nb$_{0.75}$Mo$_{0.25}$B₂ exhibits high strength of nonadiabaticity, which is quantified by the ratio of $\frac{T_\theta}{T_F} = 3.5$, which is by one order of magnitude exceeds the ratio in MgB₂, α-MoB₂, WB₂, pnictides, cuprates, and highly-compressed hydrides.

## *d*-wave superconductivity as a model for diborides apart MgB₂

## I. Introduction.

The discovery of near-room temperature superconductivity in highly compressed sulphur hydride by Drozdov *et al* [1] sparked theoretical and experimental studies of a variety of materials which potentially can exhibit a high-temperature superconductivity to be compressed at high pressure [2-25]. This research field represents one of the most fascinating scientific exploration where advanced first principles calculations conjuncts with top world class of experimental studies [26-39].

One of the interesting results in this conjunctive exploration has been reported by Pei *et al* [40] who found that the stoichiometric compound MoB₂ exhibits the phase transition from the $\beta$-MoB₂-phase (space group: $R\bar{3}m$) to $\alpha$-MoB₂-phase (space group: $P6/mmm$) at critical pressure $P \sim 70$ GPa. This high-pressure phase, $\alpha$-MoB₂, exhibits the same crystalline structure as ambient pressure MgB₂ and, what is the most intriguing experimental result reported by Pei *et al* [40], the $\alpha$-MoB₂ phase is a high-temperature superconductor with $T_c = 32\ K$ (at $P = 109.7$ GPa), which is remarkably close to $T_c = 39 - 42\ K$ in MgB₂ [41,42].

First principles calculations performed by Pei *et al* [40] showed that several bands in the $\alpha$-MoB₂ crossing the Fermi level, $\varepsilon_F$, which causes the metallic type of conductivity in this phase. Pei *et al* [40] also showed the molybdenum *d*-orbitals (especially the $d_{z2}$ orbital) have larger contributions than the boron *p*-orbitals near the $\varepsilon_F$. In overall, despite $\alpha$-MoB₂ phase exhibits the same crystal structure as MgB₂ and the superconducting transition temperature for these compounds are comparable, their electronic structures are different. For instance, the out-of-plane phonon mode of molybdenum ions are strongly coupled with molybdenum *d*-electrons near the $\varepsilon_F$ in $\alpha$-MoB₂ [40], while the in-plane boron-boron stretching mode in MgB₂ interacts intensively with the $\sigma$-bond in the boron honeycomb lattice near the $\varepsilon_F$ [40]. Pei *et al* [40] also calculated the electron-phonon coupling constant, $\lambda_{e-ph} = 1.60$, in $\alpha$-

MoB$_2$ at $P = 90\ GPa$. Similar findings, including $\lambda_{e-ph} = 1.60$, were reported by Quan *et al* [43] who performed first principles calculations for highly-pressurized $\alpha$-MoB$_2$ phase.

These results give a ground to expect that the $\alpha$-MoB$_2$ phase can exhibit *d*-wave superconducting energy gap symmetry (or, at least, *s+d*-wave gap symmetry with significant *d*-wave component), which is different from the two-band *s*-wave MgB$_2$.

More recently, Hire *et al* [44] showed that the $P6/mmm$-phase can be stabilized at ambient pressure in Nb$_{1-x}$Mo$_x$B$_2$ (x = 0.25; 0.50; 0.75 and 0.9) solid solutions. Despite the superconducting transition temperature in Nb$_{1-x}$Mo$_x$B$_2$ (x = 0.25; 0.50; 0.75 and 0.9) were significantly lower (i.e., $T_c = (6.5 - 8.1)\ K$ [44]) these values are still high enough to make a proposal that the same pairing mechanism emerges in ambient pressure superconductors Nb$_{1-x}$Mo$_x$B$_2$ and highly-pressurized $\alpha$-MoB$_2$.

Hire *et al* [44] also performed first principles calculation, measurements of the temperature dependent magnetoresistance $R(T, B)$, and specific heat measurements from which several parameters of Nb$_{1-x}$Mo$_x$B$_2$ (x = 0.25; 0.50; 0.75 and 0.9) superconductors (and, in particular, the Debye temperature, $T_\theta$) were determined.

Pei *et al* [45] and Lim *et al* [46] extended the range of superconducting diborides by the discovery of highly-compressed phase of WB$_2$ ($T_c$~15 *K* at *P*~121 GPa) for which Pei *et al* [45] proposed space group: *P6$_3$/mmc* (which is distorted *P6/mmm*), while Lim *et al* [46] concluded that this highly-pressurized superconducting phase of WB$_2$ formed by staking faulted *P6$_3$/mmc-P6/mmm* phase (which can be found to be similar to the stacking faulted 123-124 phases in Y-Ba-Cu-O system [47-49]).

Here, we performed detailed analysis of the magnetoresistance data reported by Pei *et al* [40], Hire *et al* [44], Pei *et al* [45], and showed that the *P6/mmm*-phases of Nb$_{1-x}$Mo$_x$B$_2$ (x = 0.25; 1.0) and WB$_2$ (*P*=121.3 GPa) exhibit the *d*-wave superconducting gap symmetry. We also found that ambient pressure Nb$_{1-x}$Mo$_x$B$_2$ (x = 0.25) superconductors characterized by high strength of nonadiabaticity, which can be characterized by the ratio of $\frac{T_\theta}{T_F} = 3.5$ (where $T_F$ is the Fermi temperature, which is by more than one order of magnitude exceeds the $\frac{T_\theta}{T_F}$ ratio in MgB$_2$, α-MoB$_2$, WB$_2$, pnictides, cuprates, and highly-compressed hydrides.

## II. Results

### 2.1. P6/mmm α-MoB$_2$ (P = 109.7 GPa)

#### 2.1.1. Debye temperature and the electron-phonon coupling constant

Debye temperature, $T_\theta$, can be deduced from the fit of experimentally measured temperature dependent resistance curve, $R(T)$, to the Bloch-Grüneisen (BG) equation [50,51]. In many reports, classical BG approach is advanced by introducing so-called the saturation resistance [52-57]:

$$R(T) = \frac{1}{\frac{1}{R_{sat}} + \frac{1}{R_0 + A\left(\frac{T}{T_\theta}\right)^5 \int_0^{\frac{T_\theta}{T}} \frac{x^5}{(e^x-1)(1-e^{-x})}dx}} \qquad (1)$$

where $R_{sat}$, $R_0$, $T_\theta$ and $A$ are free fitting parameters. From the deduced $T_\theta$ and measured $T_c$ (which we defined by as strict as practically possible resistance criterion of $\frac{R(T)}{R_{norm}} \to 0$, where $R_{norm}$ is the normal state resistance at the onset of the superconducting transition, see details in [56]), the electron-phonon coupling constant, $\lambda_{e-ph}$, can be calculated as unique root of advanced McMillan equation [56]:

$$T_c = \left(\frac{1}{1.45}\right) \times T_\theta \times e^{-\left(\frac{1.04(1+\lambda_{e-ph})}{\lambda_{e-ph} - \mu^*(1+0.62\lambda_{e-ph})}\right)} \times f_1 \times f_2^* \qquad (2)$$

where

$$f_1 = \left(1 + \left(\frac{\lambda_{e-ph}}{2.46(1+3.8\mu^*)}\right)^{3/2}\right)^{1/3} \qquad (3)$$

$$f_2^* = 1 + (0.0241 - 0.0735 \times \mu^*) \times \lambda_{e-ph}^2. \qquad (4)$$

where $\mu^*$ is the Coulomb pseudopotential parameter, which we assumed (follow the approach proposed in [40,44,46]) to be $\mu^* = 0.13$ for $Nb_{1-x}Mo_xB_2$ (x = 0.25; 1.0) and $WB_2$.

The fits of $R(T)$ datasets measured for $\alpha$-$MoB_2$ phase at $P = 91.4$ $and$ $109.7$ $GPa$ [40] to Eq. 1 together with deduced $R_{sat}$, $T_\theta$, and $\lambda_{e-ph}$, are shown in Fig. 1 (where we utilized $\frac{R(T)}{R_{norm}(T)} = 0.10$ criterion to define $T_c$, because the same criterion was used by Pei $et$ $al$ [40] to define the upper critical field in the same $\alpha$-$MoB_2$ sample). Deduced $\lambda_{e-ph}(91.4$ $GPa) = 1.42$ is in a good agreement with the value calculated by first principles calculations $\lambda_{e-ph}(90$ $GPa) = 1.60$ [40,43].

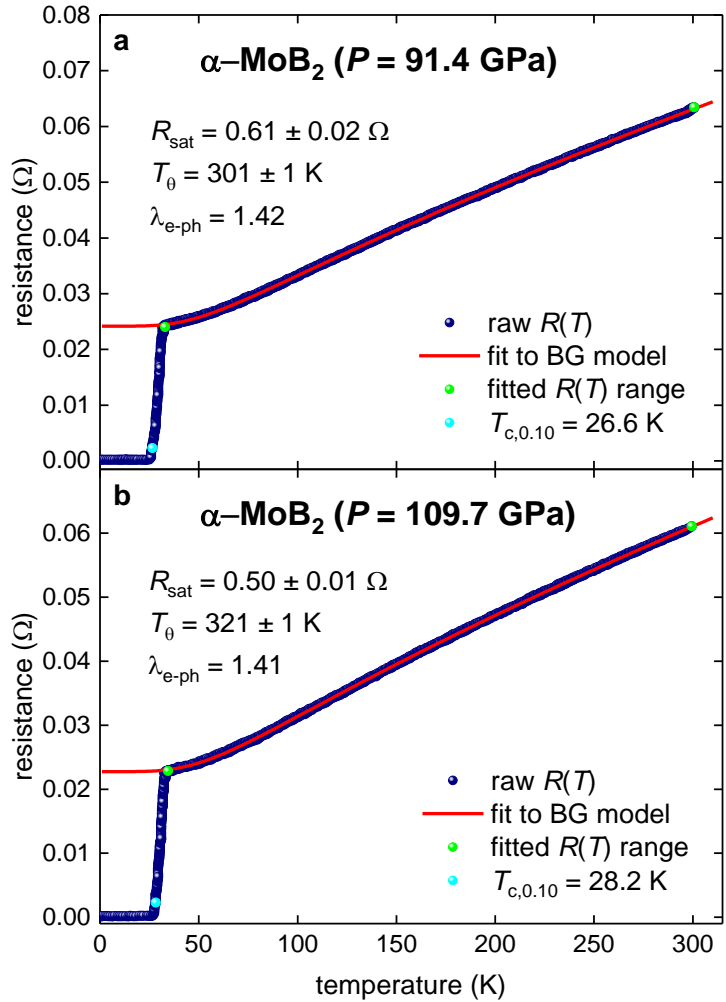

**Figure 1.** $R(T)$ data for highly compressed $\alpha$-$MoB_2$ ($P$ = 109.7 GPa) and data fit to Eq. 1 (raw data reported by Pei $et$ $al$ [40]). Green balls indicate the bounds for which $R(T)$ data was used for the fit to Eq. 1. (a) Deduced $T_\theta = 301 \pm 1$ $K$, $T_{c,0.10} = 26.6$ $K$, $\lambda_{e-ph} = 1.42$, $R_{sat} = 0.61 \pm 0.02$ $\Omega$, fit quality is 0.9998. (b) Deduced $T_\theta = 321 \pm 1$ $K$, $T_{c,0.10} = 28.2$ $K$, $\lambda_{e-ph} = 1.41$, $R_{sat} = 0.50 \pm 0.01$ $\Omega$, fit quality is 0.9998. 95% confidence bands are shown by pink shadow areas.

## 2.1.2. Temperature dependent upper critical field

Pei *et al* [40] in their Figure 2,d utilized several resistance criteria $\frac{R(T)}{R_{norm}(T)} = 0.10, 0.50, 0.90$ to derive the upper critical field, $B_{c2}(T)$, from measured $R(T, B, P = 109.7\ GPa)$ curves. By following general logic [56,58,59] that as low as possible resistance criterion should be in use, here we utilized the same criterion of $\frac{R(T)}{R_{norm}(T)} = 0.10$, as the one which was used to define the $T_c$ in Fig. 1 and by the lowest criterion to defined $B_{c2}(T)$ by Pei *et al* [40].

In Fig. 2,a the $B_{c2}(T)$ dataset is fitted to the equation for temperature dependent upper critical field for *s*-wave superconductors [58-60]:

$$B_{c2}(T) = \frac{\phi_0}{2\cdot\pi\cdot\xi^2(0)} \left(\frac{1.77 - 0.43\left(\frac{T}{T_c}\right)^2 + 0.07\left(\frac{T}{T_c}\right)^4}{1.77}\right)^2 \times \left[1 - \frac{1}{2k_B T}\int_0^\infty \frac{d\varepsilon}{\cosh^2\left(\frac{\sqrt{\varepsilon^2 + \Delta^2(T)}}{2k_B T}\right)}\right] \quad (5)$$

where the amplitude of temperature dependent superconducting gap, $\Delta(T)$, is given by [61,62]:

$$\Delta(T) = \Delta(0) \times \tanh\left[\frac{\pi k_B T_c}{\Delta(0)}\sqrt{\eta\frac{\Delta C}{\gamma T_c}\left(\frac{T_c}{T} - 1\right)}\right] \quad (6)$$

where $\eta = 2/3$ for *s*-wave superconductors, $\gamma$ is Sommerfeld constant, and $k_B$ is the Boltzmann constant. However, the deduced $\frac{2\Delta(0)}{k_B T_c} = 2.3 \pm 0.1$ (Fig. 2,a) is too low to be attributed to *s*-wave superconductivity, for which the weak-coupling limit is $\frac{2\Delta(0)}{k_B T_c} = 3.53$ [63,64]. And also, the fit quality is low $R = 0.8267$.

Then, we fitted the temperature dependent upper critical field data to the *d*-wave gap symmetry model. The fitting function can be constructed similarly to its counterparts for *s*-wave [58-60] and *p*-wave [59,60,65]:

$$B_{c2}(T) = \frac{\phi_0}{2\cdot\pi\cdot\xi^2(0)}\left(\frac{1.77 - 0.43\left(\frac{T}{T_c}\right)^2 + 0.07\left(\frac{T}{T_c}\right)^4}{1.77}\right)^2 \left[1 - \frac{1}{2\cdot k_B\cdot T}\cdot\int_0^{2\pi}\cos^2(\theta)\cdot\left(\int_0^\infty\frac{d\varepsilon}{\cosh^2\left(\frac{\sqrt{\varepsilon^2 + \Delta^2(T,\theta)}}{2\cdot k_B\cdot T}\right)}\right)\cdot d\theta\right] \quad (7)$$

where the superconducting energy gap, $\Delta(T,\theta)$, is given by [61,62,65]:

$$\Delta(T,\theta) = cos(2\theta) \times \Delta_m(0) \times \tanh\left[\frac{\pi k_B T_c}{\Delta(0)}\sqrt{\eta \frac{\Delta C}{\gamma T_c}\left(\frac{T_c}{T}-1\right)}\right] \quad (8)$$

where $\Delta_m(T)$ is the is the maximum amplitude of the $k$-dependent $d$-wave gap, $\eta = 7/5$ [65], $\theta$ is the angle around the Fermi surface subtended at $(\pi, \pi)$ in the Brillouin zone (details can be found elsewhere [61,62]).

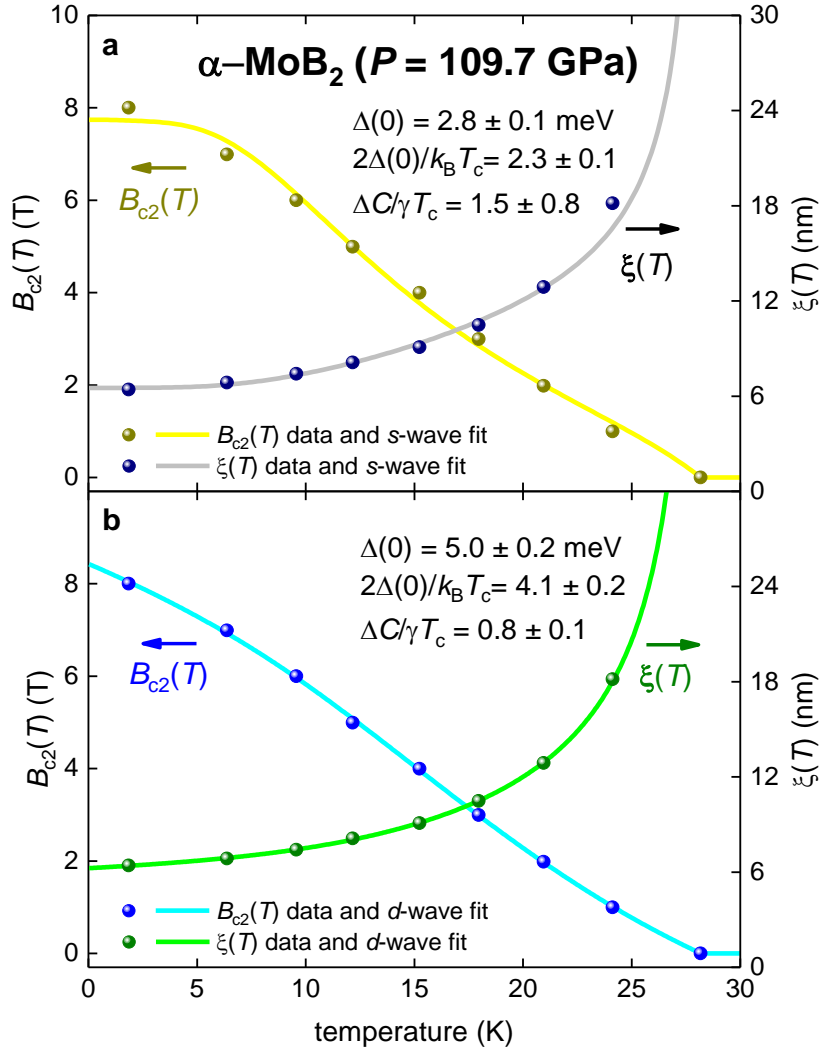

**Figure 2.** The upper critical field, $B_{c2}(T)$, data (defined by $\frac{R(T)}{R_{norm}(T)} = 0.10$ criterion) for $\alpha$-MoB$_2$ ($P = 109.7\ GPa$) reported by Pei $et\ al$ [40] and data fits to $s$-wave (panel a) and $d$-wave (panel b) single-band models. Deduced parameters are (for both panels the critical temperature was fixed to the observed value of $T_c = 28.2$ K): (a) $s$-wave fit, $\xi(0) = 6.5(2)$ nm, $\Delta(0) = 2.8 \pm 0.1\ meV$, $\Delta C/\gamma T_c = 1.5 \pm 0.8$, $\frac{2\Delta(0)}{k_B T_c} = 2.3 \pm 0.2$, the goodness of fit is 0.8267; (b) $d$-wave fit, $\xi(0) = 6.2(5)$ nm, $\Delta(0) = 5.0 \pm 0.2\ meV$, $\Delta C/\gamma T_c = 0.8 \pm 0.1$, $\frac{2\Delta(0)}{k_B T_c} = 4.1 \pm 0.2$, the goodness of fit is 0.9842.

The fit converged with a better quality (with the goodness of fit is 0.9842) (Fig. 2,b). Deduced parameters are: $\xi(0) = 6.2(5)\ nm$, $\Delta(0) = 5.0 \pm 0.2\ meV$, $\frac{2\Delta(0)}{k_B T_c} = 4.1 \pm 0.2$, $\frac{\Delta C}{\gamma T_c} = 0.8 \pm 0.1$. Considering that the weak coupling limits for *d*-wave superconductors [61,62,65] are: $\frac{2 \cdot \Delta(0)}{k_B \cdot T_c} = 4.28$ and $\frac{\Delta C}{\gamma T_c} = 0.995$, we can conclude that deduced parameters in $\alpha$-MoB$_2$ ($P = 109.7\ GPa$) superconductor is within weak-coupling values for *d*-wave superconductor.

It should be noted, that the accuracy of the extracted parameters is directly related to the sampling number of the measurement, and, thus, further increase in the accuracy in the deduced parameters, can be possible if more raw $R(T, B)$ data (especially, measured at low temperature, down to miliKelvin level) will be available.

### 2.1.3. The Fermi temperature and the strength of the nonadiabaticity

The Fermi temperature can be calculated by the equation [58]:

$$T_F = \frac{\pi^2 m_e}{8 \cdot k_B} \times (1 + \lambda_{e-ph}) \times \xi^2(0) \times \left(\frac{2\Delta_m(0)}{\hbar}\right)^2, \qquad (9)$$

where $m_e$ is bare electron mass, $\hbar$ is the reduced Planck's constant, and other parameters have deduced above. In the result, calculated Fermi temperature is $T_F = 1756 \pm 25\ K$.

Calculated $T_F$ implies that the *P6/mmm* $\alpha$-MoB$_2$ ($P = 109.7\ GPa$) phase falls in unconventional superconductors band in the Uemura plot (Fig. 3), because this phase exhibits typical for many unconventional superconductors (for instance, iron-based, cuprates and hydrogen-rich superconductors) ratio of $\frac{T_c}{T_F} = 0.016$.

Also, we found that the *P6/mmm* $\alpha$-MoB$_2$ ($P = 109.7\ GPa$) phase exhibits similar level of the nonadiabaticy ($\frac{T_\theta}{T_F} = 0.18 \pm 0.02$) to iron-based, cuprates and hydrogen-rich superconductors [66,67] (Figs. 4,5). It can be seen that Fig. 4 agrees with recent result reported by Yuzbashyan and Altshuler [75] that the electron-phonon coupling constant has the upper limit of $\lambda_{e-ph} \leq 3.7$.

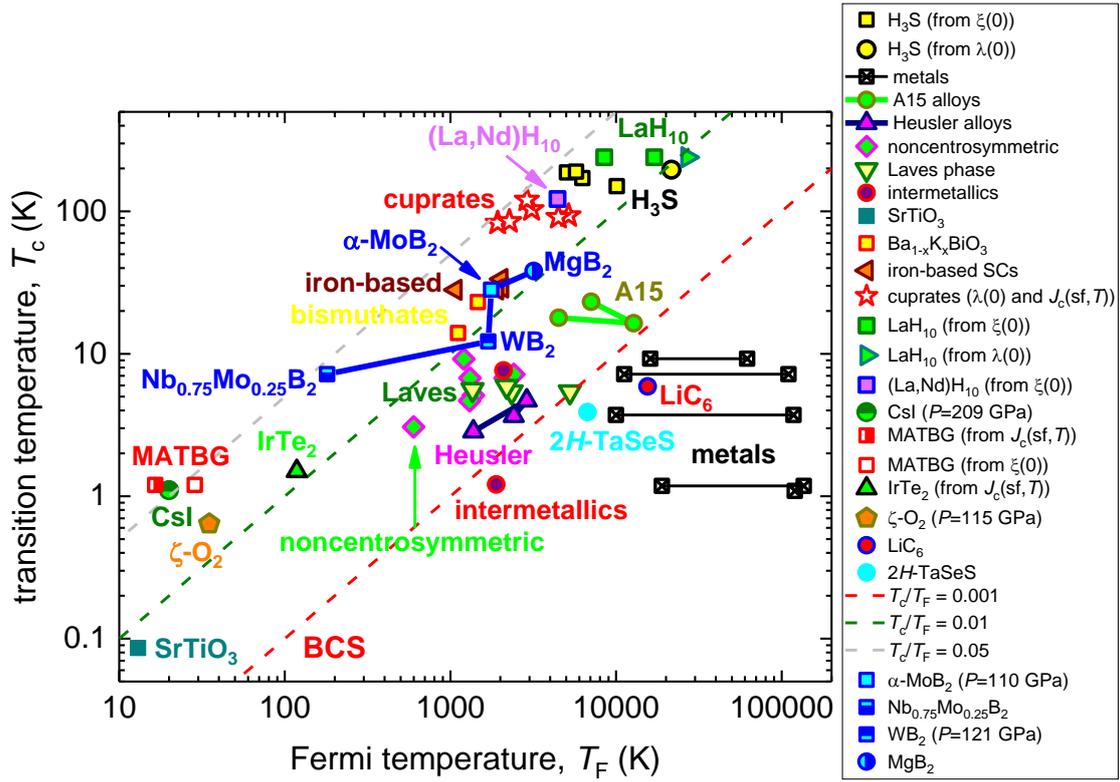

**Figure 3.** Uemura plot ($T_c$ vs $T_F$), where the diborides are shown together with other superconducting families: 2D materials, metals, pnictides, cuprates, and near-room-temperature superconductors. Reference on original data can be found in Refs. 65,67-74.

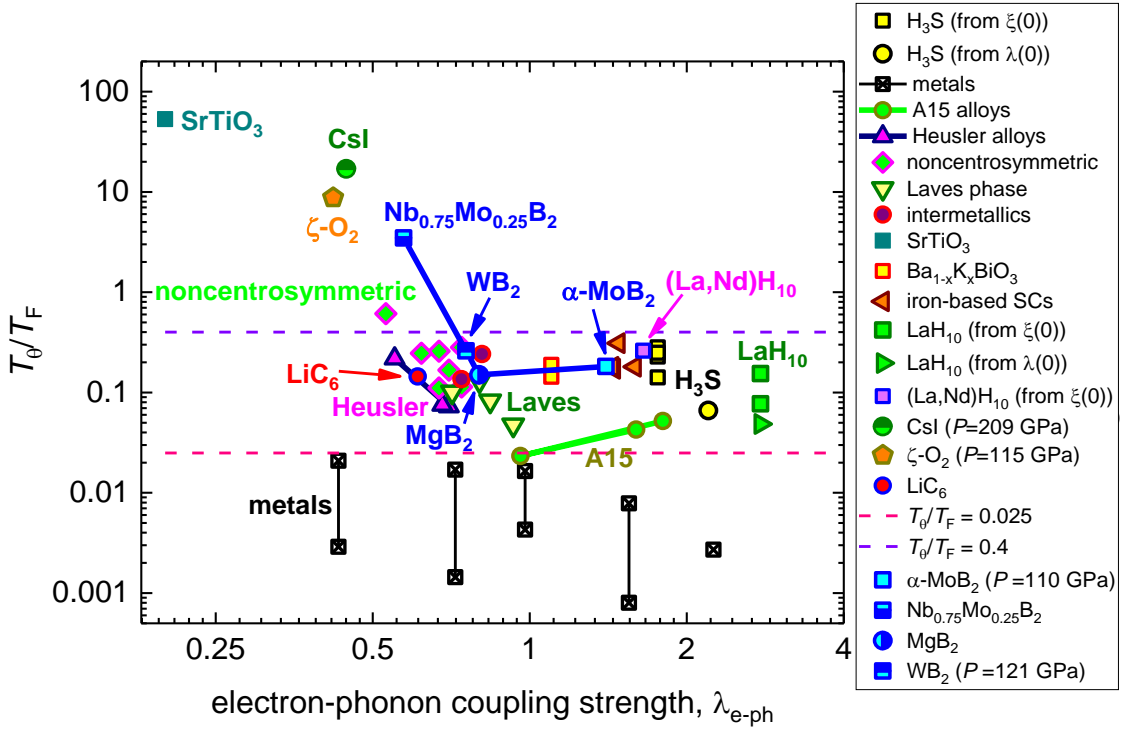

**Figure 4.** Plot of $\frac{T_\theta}{T_F}$ vs $\lambda_{e-ph}$ for several superconducting families and for diborides. This type of plot proposed by Pietronero *et al* [66]. References on original data can be found in Refs. 65,67-71.

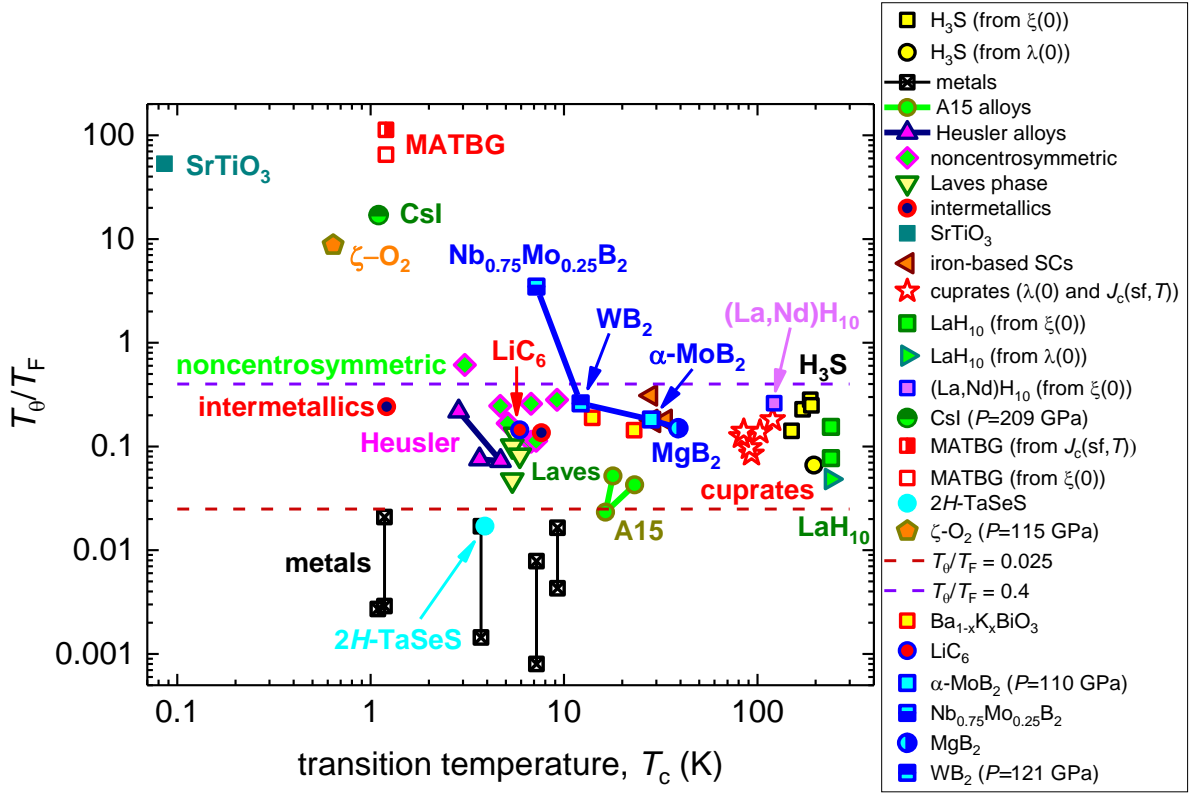

**Figure 5.** Plot of $\frac{T_\theta}{T_F}$ vs $T_c$ for several superconducting families and for the diborides. References on original data can be found in Refs. 65,67-72,74.

## 2.2. Ambient pressure P6/mmm Nd$_{0.75}$Mo$_{0.25}$B$_2$

### 2.2.1. The electron-phonon coupling constant

Hire *et al* [44] in their Table 1 reported the Debye temperature for *P6/mmm* Nb$_{1-x}$Mo$_x$B$_2$ ($x = 0.25$) which was deduced from low-temperature specific heat measurements, $T_\theta = 625\ K$. Follow the approach implemented in this study, we processed $R(T, B = 0)$ data reported by Hire *et al* [44] by utilizing the resistance criterion of $\frac{R(T)}{R_{norm}(T)} = 0.015$ and deduced $T_{c,0.015} = 7.2\ K$, from which $\lambda_{e-ph} = 0.573$ was calculated by Eqs. 2-4.

### 2.2.2. Temperature dependent upper critical field

Hire *et al* [44] in their Figure 6 reported $R(T, B)$ data reported, which we processed by utilizing the resistance criterion of $\frac{R(T)}{R_{norm}(T)} = 0.015$ and deduced the $B_{c2}(T)$ dataset. The fits of this dataset to *s*-wave (Eqs. 5,6) and *d*-wave model (Eqs. 7,8) are shown in Fig. 6.

The deduced parameters for *s*-wave (Fig. 6,a) contradict to each other, i.e $\frac{2\Delta(0)}{k_B T_c} = 3.18 \pm 0.15$ (which is lower than the *s*-wave weak-coupling limit is $\frac{2\Delta(0)}{k_B T_c} = 3.53$ [63,64]), while deduced $\frac{\Delta C}{\gamma T_c} = 1.62 \pm 0.19$ is larger than *s*-wave weak-coupling limit of $\frac{\Delta C}{\gamma T_c} = 1.43$. And the fit quality *R* =0.9534 is not high.

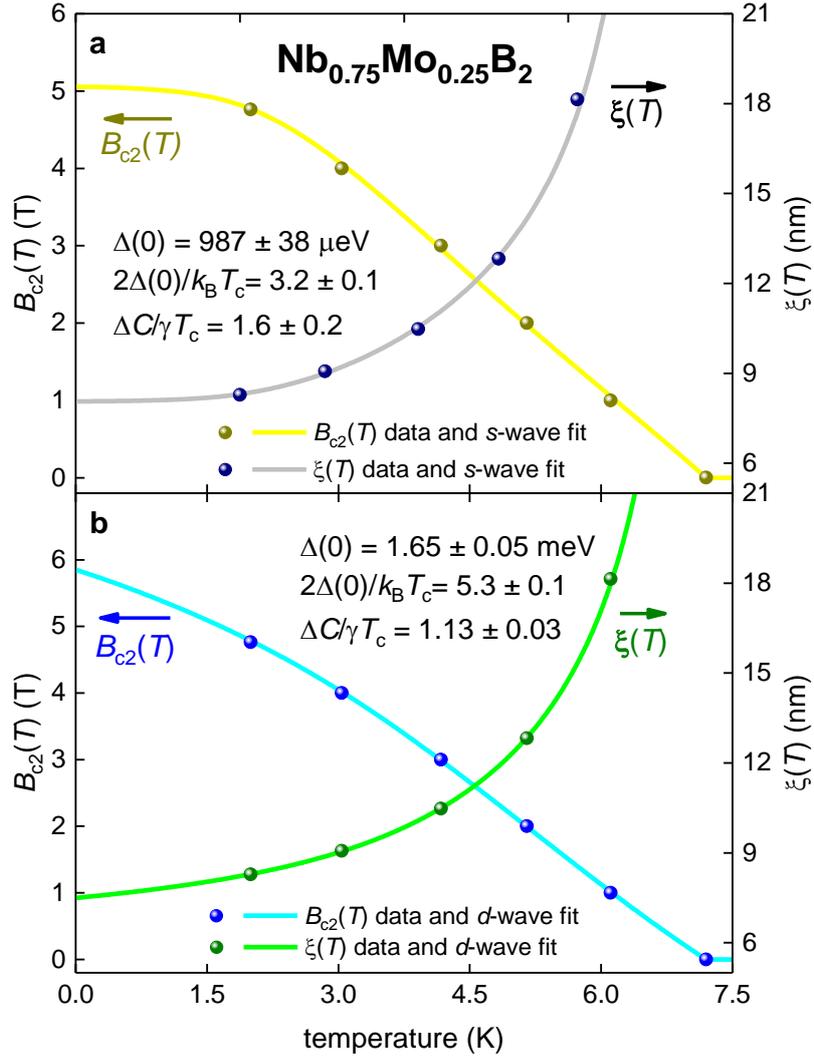

**Figure 6.** The upper critical field, $B_{c2}(T)$, data (defined by $\frac{R(T)}{R_{norm}(T)} = 0.015$ criterion) for *P6/mmm* Nb$_{0.75}$Mo$_{0.25}$B$_2$ reported by Hire *et al* [44] and data fits to *s*-wave (panel a) and *d*-wave (panel b) single-band models. Deduced parameters are (for both panels the critical temperature was fixed to the observed value of $T_c$ = 7.2 K): (a) *s*-wave fit, ξ(0) = 8.0(7) nm, $\Delta(0) = 0.987 \pm 0.038\ meV$, $\Delta C/\gamma T_c = 1.6 \pm 0.2$, $\frac{2\Delta(0)}{k_B T_c} = 3.2 \pm 0.1$, the goodness of fit is 0.9534; (b) *d*-wave fit, ξ(0) = 7.5(0) nm, $\Delta(0) = 1.65 \pm 0.05\ meV$, $\Delta C/\gamma T_c = 1.13 \pm 0.03$, $\frac{2\Delta(0)}{k_B T_c} = 5.3 \pm 0.1$, the goodness of fit is 0.9959.

The fit to the *d*-wave gap symmetry model has a better quality (with the goodness of fit is 0.9959) (Fig. 6,b). Deduced parameters are: $\xi(0) = 6.2(5)\ nm$, $\Delta(0) = 1.65 \pm 0.05\ meV$, $\frac{2\Delta(0)}{k_B T_c} = 5.3 \pm 0.1$, $\frac{\Delta C}{\gamma T_c} = 1.13 \pm 0.03$, characterize the material as moderately strong coupled *d*-wave superconductor (considering that the weak coupling limits for *d*-wave superconductors [61,62,65] are: $\frac{2 \cdot \Delta(0)}{k_B \cdot T_c} = 4.28$ and $\frac{\Delta C}{\gamma T_c} = 0.995$).

### 2.2.3. The Fermi temperature and the strength of the nonadiabaticity

The substitution of deduced parameters in Eq. 9 returns the Fermi temperature $T_F = 180 \pm 7\ K$ in *P6/mmm*-phase of Nb$_{0.75}$Mo$_{0.25}$B$_2$. Calculated $T_F$ implies that this phase falls in unconventional superconductors band in the Uemura plot (Fig. 3), because this phase exhibits typical for many unconventional superconductors ratio of $\frac{T_c}{T_F} = 0.042$.

However, what comes from our analysis and reported by Hire *et al* [44] the Debye temperature, that *P6/mmm*-phase of Nb$_{0.75}$Mo$_{0.25}$B$_2$ superconductor exhibits strong nonadiabaticy, because the ratio:

$$0.4 \ll \frac{T_\theta}{T_F} = 3.5 \pm 0.3 \tag{10}$$

is well above typical range for moderate level of nonadiabaticity ($0.025 \leq \frac{T_\theta}{T_F} \leq 0.4$) observed in majority of unconventional superconductors, including iron-based, cuprates and highly compressed hydrides [67] (Figs. 4,5).

### *2.3. P6$_3$/mmc WB$_2$ (P = 121.3 GPa)*

#### 2.3.1. The Debye temperature and the electron-phonon coupling constant

Pei *et al* [45] measured $R(T)$ datasets for WB$_2$ phase at $P = 121.3\ GPa$ which we fitted to Eq. 1 in Fig. 7. The fit converged at $T_\theta = 440 \pm 1\ K$ and $R_{sat} \to \infty$. From deduced $T_\theta$ we found $\lambda_{e-ph} = 0.755$, for which we utilized the criterion of $\frac{R(T)}{R_{norm}(T)} = 0.18$, which is based on the presence of the inflection of the transition $R(T, B, P = 121.3\ GPa)$ which can be seen in Fig. 2(b,d) of Ref. 45.

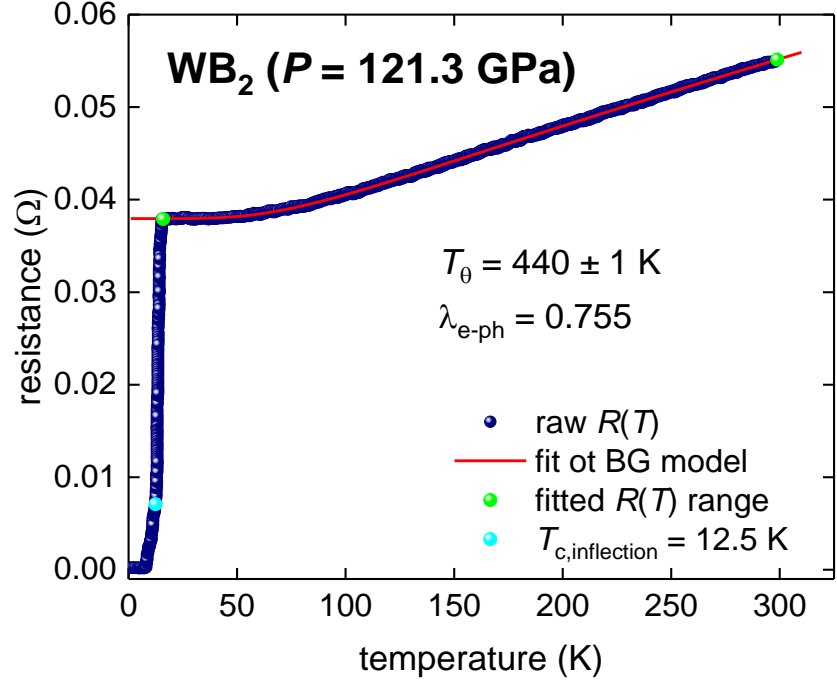

**Figure 7.** $R(T)$ data for highly compressed $WB_2$ ($P$ = 121.3 GPa) and data fit to Eq. 1 (raw data reported by Pei *et al* [45]). Green balls indicate the bounds for which $R(T)$ data was used for the fit to Eq. 1. (a) Deduced $T_\theta = 440 \pm 1\ K$, $T_{c,0.18} = 12.5\ K$, $\lambda_{e-ph} = 0.755$, $R_{sat} = \infty$, fit quality is 0.9997. 95% confidence bands are shown by pink shadow areas.

### 2.3.2. Temperature dependent upper critical field

By utilizing the resistance criterion of $\frac{R(T)}{R_{norm}(T)} = 0.18$ for $R(T, B)$ data reported by Pei *et al* [45] in their Figure 2,d we deduced the $B_{c2}(T)$ dataset for $WB_2$ ($P$ = 121.3 GPa). The fit of the $B_{c2}(T)$ dataset to *s*-wave (Eqs. 5,6) and *d*-wave model (Eqs. 7,8) are shown in Fig. 8.

The deduced parameters for *s*-wave (Fig. 6,a) contradict to each other, i.e. $\frac{2\Delta(0)}{k_B T_c} = 2.8 \pm 0.1$ (which is lower than the *s*-wave weak-coupling limit of $\frac{2\Delta(0)}{k_B T_c} = 3.53$ [63,64]), while deduced $\frac{\Delta C}{\gamma T_c} = 1.6 \pm 0.4$ is slightly larger than *s*-wave weak-coupling limit of $\frac{\Delta C}{\gamma T_c} = 1.43$. And the fit quality $R = 0.9019$ is not high.

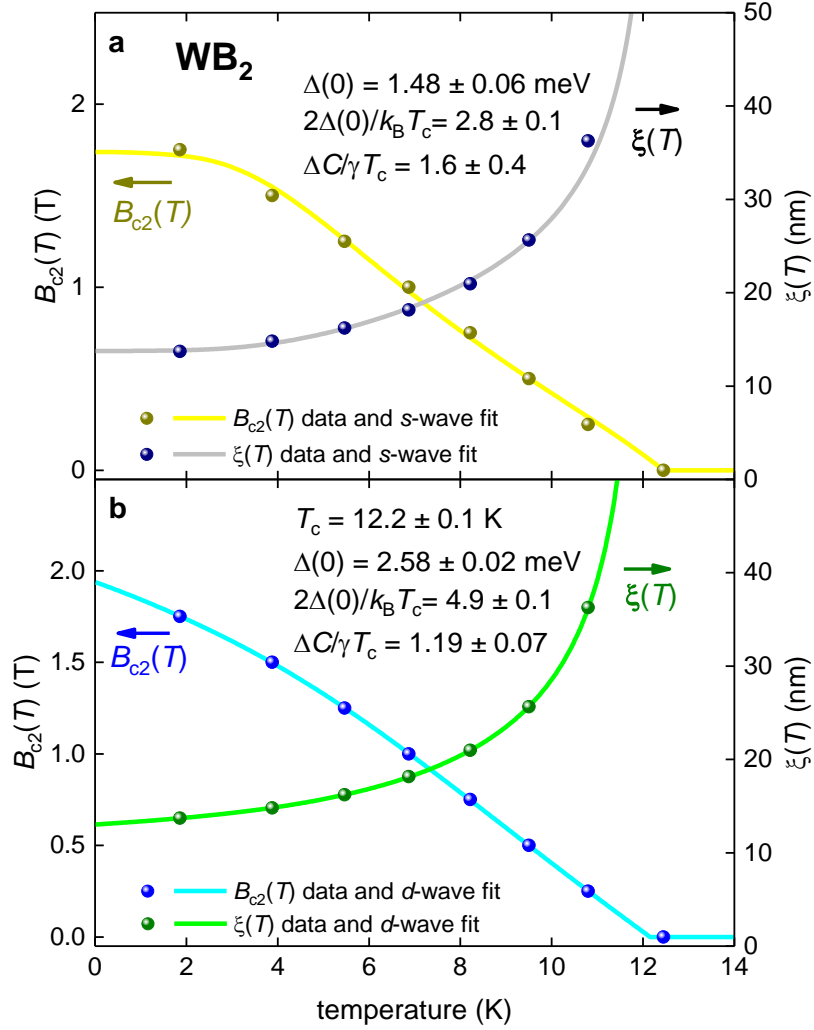

**Figure 8.** The upper critical field, $B_{c2}(T)$, data (defined by $\frac{R(T)}{R_{norm}(T)} = 0.015$ criterion) for $P6_3/mmc$ WB$_2$ ($P$ = 121.3 GPa) reported by Pei *et al* [45] and data fits to *s*-wave (panel a) and *d*-wave (panel b) single-band models. Deduced parameters are: (a) *s*-wave fit, $T_c$ = 12.45 K (fixed), ξ(0) = 13.8 nm, $\Delta(0) = 1.48 \pm 0.06\ meV$, $\Delta C/\gamma T_c = 1.6 \pm 0.4$, $\frac{2\Delta(0)}{k_B T_c} = 2.8 \pm 0.1$, the goodness of fit is 0.9019; (b) *d*-wave fit, $T_c = 12.2 \pm 0.2\ K$, ξ(0) = 13.0 nm, $\Delta(0) = 2.58 \pm 0.02\ meV$, $\Delta C/\gamma T_c = 1.19 \pm 0.07$, $\frac{2\Delta(0)}{k_B T_c} = 4.9 \pm 0.1$, the goodness of fit is 0.9986.

The fit to the *d*-wave gap symmetry model has a better quality (with the goodness of fit is 0.9986) (Fig. 8,b). Deduced parameters are: ξ(0) = 13.0 *nm*, $\Delta(0) = 2.58 \pm 0.02\ meV$, $\frac{2\Delta(0)}{k_B T_c} = 4.9 \pm 0.1$, $\frac{\Delta C}{\gamma T_c} = 1.19 \pm 0.07$, characterize the material as moderately strong coupled *d*-wave superconductor (considering that the weak coupling limits for *d*-wave superconductors [61,62,65] are: $\frac{2 \cdot \Delta(0)}{k_B \cdot T_c} = 4.28$ and $\frac{\Delta C}{\gamma T_c} = 0.995$).

### 2.3.3. The Fermi temperature and the strength of the nonadiabaticity

The substitution of deduced parameters in Eq. 9 returns the Fermi temperature $T_F = 1679 \pm 68\ K$ in $WB_2$ ($P$ = 121.3 GPa). Calculated $T_F$ implies that this phase falls in nearly conventional superconductors band in the Uemura plot (Fig. 3), because this phase exhibits reasonably low ratio of $\frac{T_c}{T_F} = 0.0077 \pm 0.0003$, while typical range for unconventional superconductors is $0.01 \leq \frac{T_c}{T_F} \leq 0.05$.

Also, this superconductor exhibits very moderate strength of nonadiabaticy, because the ratio:

$$0.025 < \frac{T_\theta}{T_F} = 0.26 \pm 0.01 < 0.4 \qquad (11)$$

is typical for majority of high-temperature superconductors, including iron-based, cuprates and highly compressed hydrides [67] (Figs. 4,5).

### *2.4. P6/mmm MgB$_2$*

### 2.4.1. Temperature dependent upper critical field

To show that our $B_{c2}(T)$ model (Eqs. 5-8 [58-60,71]) can be considered as an alternative model to extract primary superconducting parameters from $R(T,B)$ datasets (while the $B_{c2}(T)$ definition criterion is $\frac{R(T)}{R_{norm}(T)} \to 0$) in addition to widely used Werthamer-Helfand-Hohenberg model [76,77], in Figure 9 we showed $B_{c2}(T)$ data reported by Zehetmayer *et al* [78] for single crystal $MgB_2$ and datafits to the *s*-wave (panel a, Eqs. 5,6), *d*-wave (panel b, Eqs. 7,8), and so-called two-band α-model [79] in assumption of *s*-wave gap symmetry for both bands (panel c) [79,80]:

$$B_{c2,total}(T) = \alpha \times B_{c2,band1}(\xi_{total}(0), T) + (1-\alpha) \times B_{c2,band2}(\xi_{total}(0), T), \qquad (12)$$

where to reduce the number of free-fitting parameters, we implemented the restriction [80]:

$$T_{c1} = T_{c2}, \qquad (13)$$

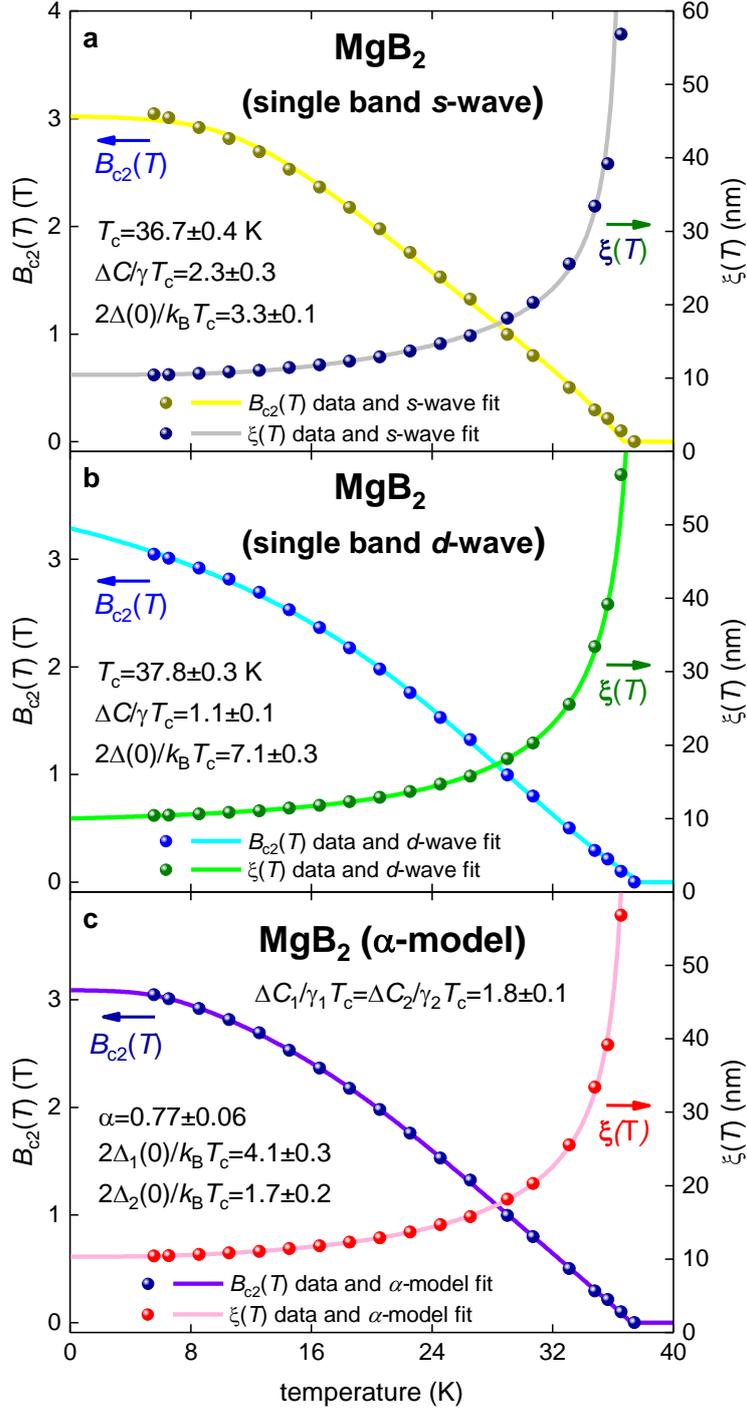

**Figure 9.** The upper critical field, $B_{c2}(T)$, data for *P6/mmm* MgB$_2$ reported by Zehetmayer *et al* [78] and data fits to single band *s*-wave (panel a, Eqs. 5,6), single band *d*-wave (panel b, Eqs. 7,8), and two-band *s*-wave [79,80] (panel c, Eqs. 5,6,12-14) models. Deduced parameters are: (a) *s*-wave fit, $T_c = 36.7 \pm 0.4\,K$, $\xi(0) = 10.4$ nm, $\Delta(0) = 5.22 \pm 0.09\,meV$, $\Delta C/\gamma T_c = 2.3 \pm 0.3$, $\frac{2\Delta(0)}{k_B T_c} = 3.3 \pm 0.1$, the goodness of fit is 0.9887; (b) *d*-wave fit, $T_c = 37.8 \pm 0.3\,K$, $\xi(0) = 10.0$ nm, $\Delta_m(0) = 11.6 \pm 0.5\,meV$, $\Delta C/\gamma T_c = 1.15 \pm 0.07$, $\frac{2\Delta_m(0)}{k_B T_c} = 7.1 \pm 0.3$, the goodness of fit is 0.9975. (c) two conditions where used: $T_{c1} = T_{c2} = 37.2 \pm 0.2\,K$ and $\frac{\Delta C_1}{\gamma_1 T_{c1}} = \frac{\Delta C_2}{\gamma_2 T_{c2}} = 1.8 \pm 0.1$, and other free-fitting parameters are: $\xi_{\text{total}}(0) = 10.3$ nm, $\alpha = 0.77 \pm 0.06$, $\Delta_1(0) = 6.5 \pm 0.4\,meV$, $\frac{2\Delta_1(0)}{k_B T_c} = 4.1 \pm 0.3$, $\Delta_2(0) = 2.7 \pm 0.4\,meV$, $\frac{2\Delta_2(0)}{k_B T_c} = 1.7 \pm 0.2$, the goodness of fit is 0.9984.

$$\frac{\Delta C_1}{\gamma_1 T_{c1}} = \frac{\Delta C_2}{\gamma_2 T_{c2}}. \tag{14}$$

The deduced parameters for single band *s*-wave model (Fig. 9,a) contradict to each other, i.e. $\frac{2\Delta(0)}{k_B T_c} = 3.3 \pm 0.1$ (which is lower than the *s*-wave weak-coupling limit), while $\frac{\Delta C}{\gamma T_c} = 2.3 \pm 0.3$ is much larger than the *s*-wave weak-coupling limit. The deduced ratio of $\frac{2\Delta_m(0)}{k_B T_c} = 7.1 \pm 0.3$ for *d*-wave model is nearly twice larger the *d*-wave weak-coupling limit of $\frac{2\Delta_m(0)}{k_B T_c} = 4.28$, which is too large to be realistic value.

However, parameters deduced for two-band α-model, $\alpha = 0.77 \pm 0.06$, $\frac{2\Delta_1(0)}{k_B T_c} = 4.1 \pm 0.3$, $\frac{2\Delta_2(0)}{k_B T_c} = 1.7 \pm 0.2$, are in a good agreement with the values deduced in MgB$_2$ by other techniques [79], in particular by point contact spectroscopy [81].

### III. Discussion

Considering that the *P6/mmm*-phase of Nb$_{0.75}$Mo$_{0.25}$B$_2$ exhibits pronounced nonadiabaticity, $\frac{T_\theta}{T_F} = 3.5$, which is well above an empirical boarder $\frac{T_\theta}{T_F} \cong 0.4$ below which the majority of conventional and unconventional superconductors are located (Figs. 4,5), we can propose that the strength of the nonadiabaticity is a primary reason for relatively low $T_c$ in this material in comparison with other diboride counterparts. A good support of this hypnotize can be seen in Fig. 5, where the $T_c$ suppression within four dibories is link with the increase of the strength of the nonadiabaticity. It can be also seen in Fig. 5, that there are no materials which simultaneously exhibite $T_c > 10\ K$ and $\frac{T_\theta}{T_F} > 0.4$.

Another explanation for the relatively low $T_c$ in Nb$_{0.75}$Mo$_{0.25}$B$_2$ can be based on the Abrikosov-Gor'kov [82], Anderson [83], and Openov [84,85] theory of dirty superconductors. The theory states that uniformly distributed (on the atomic level) impurities exhibited magnetic moment should suppress the superconducting order parameter in *s*-wave superconductors, but this kind of impurities should not affect the superconducting order parameter in *d*-wave superconductors. However, non-magnetic impurities should cause the suppression of in *d*-wave superconductors, but this kind of doping should not affect the *s*-wave superconducting state. Considering that the (Nb,Mo)-(0001) planes in *P6/mmm*-phase have chemical atomic disorder, because Hire *et al* [44] did not report any evidence for the atomic ordering within Nb-Mo atoms in (0001) planes, it is appeared that the $T_c$ suppression in $Nb_{0.75}Mo_{0.25}B_2$ (and in all materials in the $Nb_{1-x}Mo_xB_2$ (x = 0.25; 0.50; 0.75 and 0.9) system) can be interpreted as the $T_c$ suppression in *d*-wave $MoB_2$ superconductor by nonmagnetic impurities, i.e. Nb atoms.

## IV. Conclusions

In this work, we deduced primary superconducting parameters in three diborides, i.e. *P6/mmm* phases of $Nb_{1-x}Mo_xB_2$ (x = 0.25; 1.0) and $WB_2$. It was shown that these phases exhibit *d*-wave superconducting gap symmetry. We proposed that many fold suppression of the superconducting transition temperature (down to $T_c = 8\ K$) in $Nb_{0.75}Mo_{0.25}B_2$, can be related to either strong nonadiabaticity in this phase (which exhibits the ratio $\frac{T_\theta}{T_F} = 3.5$), either to the effect of the $T_c$ suppression in *d*-wave $MoB_2$ superconductors by nonmagnetic impurity (which is Nb atoms).

### Acknowledgement

The author thanks Svetoslav A. Kuzmichev (Lomonosov Moscow State University) for the discussion about dirty *s*- and *d*-wave superconductors. The author thanks financial support provided by the Ministry of Science and Higher Education of Russia (theme "Pressure" No. AAAA-A18-118020190104-3). The research funding from the Ministry of Science and Higher Education of the Russian Federation (Ural Federal University Program of Development within the Priority-2030 Program) is gratefully acknowledged.

**Data availability statement**

The data that support the findings of this study are available from the corresponding author upon reasonable request.

**Declaration of interests**

The author declares that he has no known competing financial interests or personal relationships that could have appeared to influence the work reported in this paper.

**References**

[1] A.P. Drozdov, M. I. Eremets, I. A. Troyan, V. Ksenofontov, S. I. Shylin, Conventional superconductivity at 203 kelvin at high pressures in the sulfur hydride system. *Nature* **525**, 73-76 (2015).
[2] A. P. Drozdov, *et al*. Superconductivity at 250 K in lanthanum hydride under high pressures *Nature* **569**, 528-531 (2019)
[3] M. Somayazulu, *et al*. Evidence for superconductivity above 260 K in lanthanum superhydride at megabar pressures *Phys. Rev. Lett.* **122** 027001 (2019)
[4] D. V. Semenok, *et al*. Superconductivity at 161 K in thorium hydride $ThH_{10}$: synthesis and properties *Mater. Today* **33**, 36–44 (2020)
[5] I. A. Troyan, *et al*. Anomalous high-temperature superconductivity in $YH_6$ *Adv. Mater.* **33** 2006832 (2021)
[6] P. P. Kong, *et al*. Superconductivity up to 243 K in yttrium hydrides under high pressure *Nature Communications* **12**, 5075 (2021)
[7] D. V. Semenok, *et al*. Superconductivity at 253 K in lanthanum–yttrium ternary hydrides *Materials Today* **48**, 18-28 (2021)
[8] S. Mozaffari, *et al.* Superconducting phase diagram of $H_3S$ under high magnetic fields *Nat. Commun.* **10** 2522 (2019)
[9] D. Sun, *et al.* High-temperature superconductivity on the verge of a structural instability in lanthanum superhydride *Nat. Commun.* **12** 6863 (2021)
[10] W. Chen, *et al*. Synthesis of molecular metallic barium superhydride: Pseudocubic $BaH_{12}$. *Nat. Commun.* **12**, 273 (2021).
[11] I. A. Troyan, *et al*. Anomalous high-temperature superconductivity in $YH_6$. *Adv. Mater.* **33**, 2006832 (2021).
[12] P. Kong, *et al*. Superconductivity up to 243 K in yttrium hydrides under high pressure. *Nat. Commun.* **12**, 5075 (2021).

[13] L. Ma, *et al*. High-temperature superconducting phase in clathrate calcium hydride CaH$_6$ up to 215 K at a pressure of 172 GPa. *Phys. Rev. Lett.* **128**, 167001 (2022).
[14] D. V. Semenok, *et al*. Superconductivity at 253 K in lanthanum–yttrium ternary hydrides. *Mater. Today* **48**, 18–28 (2021).
[15] D. Zhou, Superconducting praseodymium superhydrides. *Sci. Adv.* **6**, eaax6849 (2020).
[16] F. Hong, et al. Possible superconductivity at ∼70 K in tin hydride SnH$_x$ under high pressure. *Mater. Today Phys.* **22**, 100596 (2022).
[17] W. Chen, D. V. Semenok, X. Huang, H. Shu, X. Li, D. Duan, T. Cui, A. R. Oganov. High-temperature superconducting phases in cerium superhydride with a $T_c$ up to 115 K below a pressure of 1 Megabar. *Phys. Rev. Lett.* **127**, 117001 (2021).
[18] I. Osmond, *et al*. Clean-limit superconductivity in *Im3m* H$_3$S synthesized from sulfur and hydrogen donor ammonia borane. *Phys. Rev. B* **105**, L220502 (2022).
[19] D. V. Semenok, *et al*. Effect of magnetic impurities on superconductivity in LaH$_{10}$. *Adv. Mater.* **34**, 2204038 (2022).
[20] J. Bi, *et al*. Giant enhancement of superconducting critical temperature in substitutional alloy (La,Ce)H$_9$. *Nature Communications* **13**, 5952 (2022).
[21] Z. Li, et al. Superconductivity above 200 K discovered in superhydrides of calcium. *Nat. Commun.* **13**, 2863 (2022).
[22] V. S. Minkov, *et al*. Magnetic field screening in hydrogen-rich high-temperature superconductors. *Nature Communications* **13**, 3194 (2022).
[23] V. S. Minkov, V. B. Prakapenka, E. Greenberg, M. I. Eremets. A boosted critical temperature of 166 K in superconducting D$_3$S synthesized from elemental sulfur and hydrogen. *Angewandte Chemie International Edition* **59**, 18970-18974 (2020)
[24] Z. Y. Liu, et al. Pressure-induced superconductivity up to 9 K in the quasi-one-dimensional KMn$_6$Bi$_5$. *Physical Review Letters* **128**, 187001 (2022).
[25] H. Wang, J. S. Tse, K. Tanaka, T. Iitaka, Y. Ma. Superconductive sodalite-like clathrate calcium hydride at high pressures. *PNAS* **109**, 6463-6466 (2012).
[26] A. O. Lyakhov, A. R. Oganov, H. T. Stokes, and Q. Zhu, New developments in evolutionary structure prediction algorithm USPEX, *Comput. Phys. Commun.* **184**, 1172-1182 (2013).
[27] Y. Li, J. Hao, H. Liu, Y. Li, and Y. Ma. The metallization and superconductivity of dense hydrogen sulphide. J. Chem. Phys. **140**, 174712 (2014).
[28] D. Duan, Y. Liu, F. Tian, D. Li, X. Huang, Z. Zhao, H. Yu, B. Liu, W. Tian, and T. Cui. Pressure-induced metallization of dense (H$_2$S)$_2$H$_2$ with high-$T_c$ superconductivity. *Sci. Rep.* **4**, 6968 (2014).
[29] C. Heil, S. di Cataldo, G. B. Bachelet, and L. Boeri. Superconductivity in sodalite-like yttrium hydride clathrates. *Phys. Rev. B* **99**, 220502(R) (2019).
[30] J. A. Flores-Livas, L. Boeri, A. Sanna, G. Profeta, R. Arita, and M. Eremets, A perspective on conventional high-temperature superconductors at high pressure: Methods and materials. *Phys. Rep*. 856, 1 (2020).
[31] V. Struzhkin, B. Li, C. Ji, X.-J. Chen, V. Prakapenka, E. Greenberg, I. Troyan, A. Gavriliuk, and H.-k. Mao, Superconductivity in La and Y hydrides: Remaining questions to experiment and theory. *Matter Radiat. Extrem*. **5**, 028201 (2020).
[32] I. Errea, et al. Quantum crystal structure in the 250-kelvin superconducting lanthanum hydride. *Nature* **578**, 66–69 (2020).
[33] E. Gregoryanz, C. Ji, P. Dalladay-Simpson, B. Li, R. T. Howie, and H.-K. Mao, Everything you always wanted to know about metallic hydrogen but were afraid to ask. *Matter and Radiation at Extremes* **5**, 038101 (2020)
[34] L. Boeri, *et al*. The 2021 room-temperature superconductivity roadmap *Journal of Physics: Condensed Matter* **34**, 183002 (2021).

[35] M. Shao, W. Chen, K. Zhang, X. Huang, and T. Cui. High-pressure synthesis of superconducting clathratelike YH$_4$. *Phys. Rev. B* **104**, 174509 (2021).

[36] G. Kafle, E. Komleva, Y. Amiel, A. Palevski, E. Greenberg, S. Chariton, Y. Ponosov, D. Khomskii, G. Rozenberg, S. Streltsov, I. Mazin, E. Margine. Investigation of superconducting properties of AuAgTe$_4$ under pressure. *APS March Meeting 2022.* (Vol. 67, No. 3, March 14–18, 2022; Chicago, USA) W57.00012: Investigation of superconducting properties of AuAgTe$_4$ under pressure.

[37] H.-K. Mao. Hydrogen and related matter in the pressure dimension. *Matter and Radiation at Extremes* 7, 063001 (2022).

[38] Z. Zhang, et al. Design principles for high-temperature superconductors with a hydrogen-based alloy backbone at moderate pressure. *Phys. Rev. Lett.* **128**, 047001 (2022).

[39] M. Du, H. Song, Z. Zhang, D. Duan, T. Cui. Room-temperature superconductivity in Yb/Lu substituted clathrate hexahydrides under moderate pressure. *RESEARCH* **2022**, 9784309 (2022).

[40] C. Pei, *et al*. Pressure-induced Superconductivity at 32 K in MoB$_2$. *arXiv* 2105.13250 (2021).

[41] J. Nagamatsu, N. Nakagawa, T. Muranaka, Y. Zenitani, J. Akimitsu. Superconductivity at 39 K in magnesium diboride. *Nature* **410**, 63–64 (2001).

[42] C. G. Zhuang, S. Meng, C. Y. Zhang, Q. R. Feng, Z. Z. Gan, H. Yang, Y. Jia, H. H. Wen, and X. X. Xi, Ultrahigh current-carrying capability in clean MgB$_2$ films. *J. Appl. Phys.* **104**, 013924 (2008).

[43] Y. Quan, K.-W. Lee, W. E. Pickett. MoB$_2$ under pressure: Superconducting Mo enhanced by boron. *Phys. Rev. B* **104**, 224504 (2021).

[44] A. C. Hire, *et al*. High critical field superconductivity at ambient pressure in MoB$_2$ stabilized in the *P6/mmm* structure via Nb substitution. *arXiv* 2212.14869 (2022).

[45] C. Pei, *et al*. Distinct superconducting behaviours of pressurized WB$_2$ and ReB$_2$ with different local B layers. *Sci. China-Phys. Mech. Astron*. **65**, 287412 (2022).

[46] J. Lim, A. C. Hire, Y. Quan, *et al.* Creating superconductivity in WB$_2$ through pressure-induced metastable planar defects. *Nat Commun* **13**, 7901 (2022).

[47] H. W. Zandbergen, R. Gronsky, K. Wang, and G. Thomas. Structure of (CuO)$_2$ double layers in superconducting YBa$_2$Cu$_3$O$_7$. *Nature* **331**, 596-599 (1988).

[48] E. D. Specht, A. Goyal, J. Li, P. M. Martin, X. Li, and M. W. Rupich, Stacking faults in YBa$_2$Cu$_3$O$_{7-x}$: Measurement using x-ray diffraction and effects on critical current. *Appl. Phys. Lett.* **89**, 162510 (2006).

[49] E. F. Talantsev, N. M. Strickland, S. C. Wimbush, J. G. Storey, J. L. Tallon, and N. J. Long. Hole doping dependence of critical current density in YBa$_2$Cu$_3$O$_{7-\delta}$ conductors. *Appl. Phys. Lett.* **104**, 242601 (2014).

[50] F. Bloch, Zum elektrischen Widerstandsgesetz bei tiefen Temperaturen *Z. Phys.* **59**, 208-214 (1930)

[51] E. Grüneisen, Die abhängigkeit des elektrischen widerstandes reiner metalle von der temperatur. *Ann. Phys.* **408**, 530–540 (1933)

[52] Z. Fisk and G. W. Webb, Saturation of the high-temperature normal-state electrical resistivity of superconductors, *Phys. Rev. Lett.* **36**, 1084-1086 (1976)

[53] H. Wiesmann, *et al*. Simple model for characterizing the electrical resistivity in A-15 superconductors, *Phys. Rev. Lett.* **38**, 782-785 (1977)

[54] R. Matsumoto, *et al*.Pressure-induced superconductivity in tin sulfide. *Phys. Rev. B* **99**, 184502 (2019).

[55] K. Kudo, H. Hiiragi, T. Honda, K. Fujimura, H. Idei, M. Nohara. Superconductivity in Mg$_2$Ir$_3$Si: A fully ordered Laves phase. *J. Phys. Soc. Jpn*. **89**, 013701 (2020).

[56] E. F. Talantsev, Advanced McMillan's equation and its application for the analysis of highly-compressed superconductors. *Superconductor Science and Technology* **33**, 094009 (2020).
[57] E. F. Talantsev, The electron–phonon coupling constant and the Debye temperature in polyhydrides of thorium, hexadeuteride of yttrium, and metallic hydrogen phase III. *Journal of Applied Physics* **130**, 130, 195901 (2021).
[58] E. F. Talantsev, Classifying superconductivity in compressed $H_3S$. *Modern Physics Letters B* **33**, 1950195 (2019)
[59] E. F. Talantsev, In-plane *p*-wave coherence length in iron-based superconductors. *Results in Physics* **18**, 103339 (2020).
[60] E. F. Talantsev, R. C. Mataira, W. P. Crump, Classifying superconductivity in Moiré graphene superlattices. *Scientific Reports* **10**, 212 (2020)
[61] F. Gross, *et al.* Anomalous temperature dependence of the magnetic field penetration depth in superconducting $UBe_{13}$. *Z. Phys. B* **64**, 175-188 (1986)
[62] F. Gross-Alltag, B. S. Chandrasekhar, D. Einzel, P. J. Hirschfeld and K. Andres, London field penetration in heavy fermion superconductors. *Z. Phys. B* **82**, 243-255 (1991)
[63] J. Bardeen, L. N. Cooper, J. R. Schrieffer, Theory of superconductivity. *Phys. Rev.* **108**, 1175-1204 (1957).
[64] J. P. Carbotte, Properties of boson-exchange superconductors. *Rev. Mod. Phys.* **62**, 1027-1157 (1990)
[65] E. F. Talantsev, W. P. Crump, J. L. Tallon, Thermodynamic parameters of single- or multi-band superconductors derived from self-field critical currents. *Annalen der Physik (Berlin)* **529**, 1700197 (2017)
[66] L. Pietronero, L. Boeri, E. Cappelluti, L. Ortenzi. Conventional/unconventional superconductivity in high-pressure hydrides and beyond: Insights from theory and perspectives. *Quantum Stud. Math. Found.* **5**, 5-21 (2018).
[67] E. F. Talantsev. Quantifying nonadiabaticity in major families of superconductors. *Nanomaterials* **13**, 71 (2023).
[68] E. F. Talantsev, J. L. Tallon. Universal self-field critical current for thin-film superconductors. *Nature Communications* **6**, 7820 (2015).
[69] E. Cappelluti, S. Ciuchi, C. Grimaldi, L. Pietronero, and S. Strässler. High $T_c$ superconductivity in $MgB_2$ by nonadiabatic pairing. *Physical Review Letters* **88**, 117003 (2002).
[70] K. Papagelis, *et al*. μSR studies of superconducting $MgB_{1.96}C_{0.04}$. *Physica B* **326**, 346-349 (2003).
[71] E. F. Talantsev, Electron–phonon coupling constant and BCS ratios in $LaH_{10–y}$ doped with magnetic rare-earth element. *Superconductor Science and Technology* **35**, 095008 (2022).
[72] C. Patra, T. Agarwal, Rajeshwari R. Chaudhari, and R. P. Singh. Two-dimensional multigap superconductivity in bulk 2*H*-TaSeS. *Physical Review B* **106**, 134515 (2022).
[73] S. Park, *et al*. Superconductivity emerging from a stripe charge order in $IrTe_2$ nanoflakes. *Nature Communications* **12**, 3157 (2021).
[74] V. S. Minkov, V. Ksenofontov, S. L. Budko, E. F. Talantsev, M. I. Eremets. Trapped magnetic flux in hydrogen-rich high-temperature superconductors. *arXiv*:2206.14108 (2022). 10.48550/arXiv.2206.14108.
[75] E. A. Yuzbashyan, B. L. Altshuler. Breakdown of the Migdal-Eliashberg theory and a theory of lattice-fermionic superfluidity. *Phys. Rev. B* **106**, 054518 (2022).
[76] E. Helfand and N. R. Werthamer, Temperature and purity dependence of the superconducting critical field, $H_{c2}$. II. *Phys. Rev.* **147** 288-294 (1966).

[77]  N. R. Werthamer, E. Helfand and P. C. Hohenberg, Temperature and purity dependence of the superconducting critical field, $H_{c2}$. III. Electron spin and spin-orbit effects *Phys. Rev.* **147**, 295-302 (1966).

[78]  M. Zehetmayer, M. Eisterer, J. Jun, S. M. Kazakov, J. Karpinski, A. Wisniewski, and H. W. Weber. Mixed-state properties of superconducting $MgB_2$ single crystals. *Phys. Rev. B* **66**, 052505 (2002)

[79]  F. Bouquet, Y. Wang, R. A. Fisher, D. G. Hinks, J. D. Jorgensen, A. Junod, and N. E. Phillips. Phenomenological two-gap model for the specific heat of $MgB_2$. *EPL* **56**, 856 (2001).

[80]  E. F. Talantsev, W. P. Crump, J. O. Island, Y. Xing, Y. Sun, J. Wang, J. L. Tallon. On the origin of critical temperature enhancement in atomically thin superconductors. *2D Mater.* **4**, 025072 (2017).

[81]  P. Szabó, P. Samuely, J. Kačmarčík, T. Klein, J. Marcus, D. Fruchart, S. Miraglia, C. Marcenat, and A. G. M. Jansen. Evidence for two superconducting energy gaps in $MgB_2$ by point-contact spectroscopy. *Physical Review Letters* **87**, 137005 (2001).

[82]  L. P. Gor'kov, in Superconductivity: Conventional and unconventional superconductors (Eds. K. H. Bennemann & John B. Ketterson) (Springer Berlin Heidelberg, 2008) pp. 201-224.

[83]  P. W. Anderson, Theory of dirty superconductors. *J. Phys. Chem. Solids* **11**, 26-30 (1959).

[84]  L. A. Openov, Critical temperature of an anisotropic superconductor containing both nonmagnetic and magnetic impurities *Phys Rev B* **58**, 9468-9478 (1998).

[85]  L. A. Openov, Effect of nonmagnetic and magnetic impurities on the specific heat jump in anisotropic superconductors *Phys Rev B* **69**, 224516 (2004).